\font\bbb=msbm10 \font\bbs=msbm7                                   

\input amssym.tex

\overfullrule=0pt

\def\C{\hbox{\bbb C}}

\def\Z{\hbox{\bbb Z}} \def\sZ{\hbox{\bbs Z}} 

\def\Alg{{\sl Algorithmica}}

\def\BAMS{{\sl Bull.\ Amer.\ Math.\ Soc.}}

\def\BSM{{\sl Bull.\ des Sciences Mathematiques}}

\def\CACM{{\sl Commun.\ ACM\/}}

\def\FoCM{{\sl Found.\ Comput.\ Math.}}
\def\ForP{{\sl Fortsch.\ Phys.}}

\def\IC{{\sl Inform.\ Control}}

\def\IEEETIT{{\sl IEEE Trans.\ Inform.\ Theory}}

\def\JMA{{\sl J. Multiv.\ Anal.}}

\def\ML{{\sl Machine Learning}}

\def\PIEEE{{\sl Proc.\ IEEE\/}}

\def\PM{{\sl Phil.\ Mag.}}

\def\PRL{{\sl Phys.\ Rev.\ Lett.}}
\def\PRSLA{{\sl Proc.\ Roy.\ Soc.\ Lond.\ A\/}}

\def\SIAMJC{{\sl SIAM J. Comput.}}

\def\St{{\sl Stochastics\/}}

\def\dajm{\hbox{D. A. Meyer}}

\def\bv{\hbox{E. Bernstein and U. Vazirani}}

\def\grover{\hbox{L. K. Grover}}

\def\shor{\hbox{P. W. Shor}}

\def\vonneumann{\hbox{J. von Neumann}}

\def\hfb{\hfil\break}

\catcode`@=11
\newskip\ttglue

   \font\ninerm=cmr9    \font\eightrm=cmr8   \font\sixrm=cmr6
  \font\ninebf=cmbx9   \font\eightbf=cmbx8  \font\sixbf=cmbx6
  \font\nineit=cmti9   \font\eightit=cmti8  
  \font\ninesl=cmsl9   \font\eightsl=cmsl8  
  \font\ninemi=cmmi9   \font\eightmi=cmmi8  \font\sixmi=cmmi6

\font\bigten=cmr10 scaled\magstep2 

\def\ninepoint{\def\rm{\fam0\ninerm}%
  \textfont0=\ninerm \scriptfont0=\sixrm
  \textfont1=\ninemi \scriptfont1=\sixmi
  \textfont\itfam=\nineit  \def\it{\fam\itfam\nineit}%
  \textfont\slfam=\ninesl  \def\sl{\fam\slfam\ninesl}%
  \textfont\bffam=\ninebf  \scriptfont\bffam=\sixbf
    \def\bf{\fam\bffam\ninebf}%
  \tt \ttglue=.5em plus.25em minus.15em
  \normalbaselineskip=11pt
  \setbox\strutbox=\hbox{\vrule height8pt depth3pt width0pt}%
  \normalbaselines\rm}

\def\eightpoint{\def\rm{\fam0\eightrm}%
  \textfont0=\eightrm \scriptfont0=\sixrm
  \textfont1=\eightmi \scriptfont1=\sixmi
  \textfont\itfam=\eightit  \def\it{\fam\itfam\eightit}%
  \textfont\slfam=\eightsl  \def\sl{\fam\slfam\eightsl}%
  \textfont\bffam=\eightbf  \scriptfont\bffam=\sixbf
    \def\bf{\fam\bffam\eightbf}%
  \tt \ttglue=.5em plus.25em minus.15em
  \normalbaselineskip=9pt
  \setbox\strutbox=\hbox{\vrule height7pt depth2pt width0pt}%
  \normalbaselines\rm}

\def\sfootnote#1{\edef\@sf{\spacefactor\the\spacefactor}#1\@sf
      \insert\footins\bgroup\eightpoint
      \interlinepenalty100 \let\par=\endgraf
        \leftskip=0pt \rightskip=0pt
        \splittopskip=10pt plus 1pt minus 1pt \floatingpenalty=20000
        \parskip=0pt\smallskip\item{#1}\bgroup\strut\aftergroup\@foot\let\next}
\skip\footins=12pt plus 2pt minus 2pt
\dimen\footins=30pc

\def\ie{{\it i.e.}}
\def\eg{{\it e.g.}}

\def\Remark{R{\eightpoint EMARK}}
\def\Theorem{T{\eightpoint HEOREM}}

\def\Corollary{C{\eightpoint OROLLARY}}
\def\Definition{D{\eightpoint EFINITION}}
\def\Conjecture{C{\eightpoint ONJECTURE}}
\def\Proposition{P{\eightpoint ROPOSITION}}
\def\endproof{\vrule height6pt width4pt depth2pt}

\def\and{{\eightpoint AND}}

\def\frac#1#2{{{#1}\over{#2}}}
\def\majority{M{\eightpoint AJORITY}}
\def\smallship{S{\eightpoint MALLSHIP}($d$)}
\def\bigship{B{\eightpoint IGSHIP}($\alpha$)}
\def\battleship{B{\eightpoint ATTLESHIP}}

\def\Shor{1}
\def\Coppersmith{2}
\def\Hoyer{3}
\def\Klappenecker{4}
\def\KlappeneckerRoetteler{5}
\def\Freedman{6}
\def\Grover{7}
\def\BrassardHoyer{8}
\def\GroverB{9}
\def\BHT{10}
\def\BHMT{11}
\def\BernsteinVazirani{12}
\def\machinelearning{13}
\def\Angluinsurvey{14}
\def\Angluin{15}
\def\GareyJohnson{16}
\def\noe{17}
\def\CEMM{18}
\def\Sylvester{19}
\def\Hadamard{20}
\def\vanDam{21}
\def\Helstrom{22}
\def\Kholevo{23}
\def\vonNeumann{24}
\def\Belavkin{25}
\def\Kennedy{26}
\def\Helstrombook{27}
\def\YKLa{28}
\def\procrustean{29}
\def\EldarForney{30}
\def\ServedioGortler{31}
\def\BBHT{32}
\def\BshoutyJackson{33}
\def\Valiant{34}

\magnification=1200

\dimen0=\hsize \divide\dimen0 by 13 \dimendef\chasm=0
\dimen1=\chasm \multiply\dimen1 by  6 \dimendef\halfwidth=1
\dimen2=\chasm \multiply\dimen2 by  7 \dimendef\secondstart=2
\dimen3=\chasm \divide\dimen3 by 2 \dimendef\quarter=3
\dimen4=\quarter \multiply\dimen4 by 9 \dimendef\twopointtwofivein=4
\dimen5=\chasm \multiply\dimen5 by 3 \dimendef\onepointfivein=5
\dimen6=\chasm \multiply\dimen6 by 7 \dimendef\threepointfivein=6
\dimen7=\hsize \advance\dimen7 by -\chasm \dimendef\usewidth=7
\dimen8=\chasm \multiply\dimen8 by 4 \dimendef\thirdwidth=8
\dimen9=\usewidth \divide\dimen9 by 2 \dimendef\halfwidth=9
\dimen10=\usewidth \divide\dimen10 by 3 
                   \multiply\dimen10 by 2 \dimendef\twothirdswidth=10

\parskip=0pt\parindent=0pt

\line{\hfil 1 August 2003}
\line{\hfill \tt quant-ph/0309059}
\vfill
\centerline{\bf\bigten THE GEOMETRY OF QUANTUM LEARNING}
\bigskip\bigskip
\centerline{\bf Markus Hunziker$^{*\dagger}$, David A. Meyer$^*$, 
                Jihun Park$^{\dagger\ddagger}$,}
\centerline{\bf James Pommersheim$^*$ and
                Mitch Rothstein$^{\dagger}$}
\bigskip 
\centerline{\sl {}$^*$Project in Geometry and Physics,
                Department of Mathematics}
\centerline{\sl University of California/San Diego,
                La Jolla, CA 92093-0112}
\smallskip
\centerline{\sl {}$^{\dagger}$Department of Mathematics}
\centerline{\sl University of Georgia, Athens, GA 30602-7403}
\smallskip
\centerline{\sl {}$^{\ddagger}$Department of Mathematics}
\centerline{\sl Pohang University of Science and Technology,
                Pohang, Kyungbuk 790-784, Korea}
\smallskip
\centerline{{\tt hunziker@math.uga.edu}, {\tt dmeyer@math.ucsd.edu},
            {\tt wlog@postech.ac.kr},}
\centerline{{\tt jamie@math.ucsd.edu} and {\tt rothstei@math.uga.edu}}
            
\smallskip

\vfill
\centerline{ABSTRACT}
\bigskip
\noindent Concept learning provides a natural framework in which to 
place the problems solved by the quantum algorithms of 
Bernstein-Vazirani and Grover.  By combining the tools used in these
algorithms---quantum fast transforms and amplitude 
amplification---with a novel (in this context) tool---a solution 
method for geometrical optimization problems---we derive a general 
technique for quantum concept learning.  We name this technique 
``Amplified Impatient Learning'' and apply it to construct quantum 
algorithms solving two new problems:  \battleship\ and \majority, more 
efficiently than is possible classically.

\bigskip\bigskip
\noindent 2003 Physics and Astronomy Classification Scheme:
                   02.67.Lx. 
                   
\noindent 2000 American Mathematical Society Subject Classification:
                   81P68,    
                   68Q32,    
                   15A60.    

\smallskip
\global\setbox1=\hbox{Key Words:\enspace}
\parindent=\wd1
\item{Key Words:} quantum algorithms, Procrustes problem. 

\vfill
\eject

\headline{\ninepoint\it Geometry of quantum learning 
               \hfill Hunziker, Meyer, Park, Pommersheim \& Rothstein}

\parskip=10pt
\parindent=20pt

\noindent{\bf 1.  Introduction}

Over the past decade increasing numbers of scientists have built 
quantum computation into an imposing edifice.  The paucity of quantum
algorithms, however, betrays a certain emptiness at its center.  Only
a handful of problems are known to be solvable more efficiently 
quantum mechanically than classically, and even fewer general 
quantum algorithmic techniques are known.  The latter include quantum 
fast transforms [\Shor--\Freedman]
and amplitude amplification [\Grover--\BHMT].
In this paper we 
explain how to combine these techniques with a new (in this context) 
one---a solution method for geometrical optimization problems---into 
quantum algorithms that solve new classes of problems.

These new problems can be thought of as generalizations of the 
structured and unstructured search problems solved by the quantum
algorithms of Bernstein and Vazirani [\BernsteinVazirani] and Grover
[\Grover].  Our thinking, however, is largely informed by a branch of
classical artificial intelligence---machine learning 
[\machinelearning], or more specifically, computational learning 
theory [\Angluinsurvey].

In this subject, a {\sl concept\/} is a map $c : X \to \Z_2$, defined
on some discrete set $X$; the support of the function, 
$c^{-1}(1) \subset X$, is the {\sl extension\/} of the concept.  For
example, let $X$ be the set of all balloons, and define $c(x) = 1$ if
and only if $x\in X$ is red; this concept is ``red balloon''.  
{\sl Concept learning\/} is the process by which a {\sl student\/} 
(the learner) identifies (or approximates) a target concept from a 
{\sl concept class\/} ${\cal C}$ of possible concepts.  Learning can 
be {\sl passive}---in situations where examples $x \in X$ are 
presented to the student by some external mechanism, or 
{\sl active}---in situations where the student can query a 
{\sl teacher\/} for information about the target concept.  In the 
latter case, Angluin has defined a {\sl minimally adequate teacher\/}
to consist of a {\sl pair\/} of oracles:  a {\sl membership oracle\/}
that responds to a query $x\in X$ with $\bar c(x)$, where 
$\bar c\in{\cal C}$ is the target concept; and an {\sl equivalence 
oracle\/} that responds to a query $c \in {\cal C}$ with 
$\delta_{c\bar c}$ [\Angluin].

The number of queries made by a learning algorithm is the {\sl query
complexity\/} of the algorithm; the number of queries to the 
membership oracle is its {\sl sample complexity}.  These are distinct 
from the {\sl computational complexity\/} of the algorithm, which is 
defined in the usual way [\GareyJohnson].  A {\sl family\/} of concept 
classes ${\cal C}_i$, for $0 < i \in \Z$, is an infinite sequence of 
concept classes defined on a corresponding sequence of sets $X_i$.  A 
learning algorithm for such a family is a sequence of learning 
algorithms, one for each ${\cal C}_i$.  Since each algorithm in the 
sequence has a sample complexity, we can discuss the asymptotic sample 
complexity of the family.  As we describe in detail in \S2, both 
Bernstein and Vazirani's and Grover's algorithms can be interpreted as 
quantum algorithms for concept learning from a membership oracle, each 
with a sample complexity that is asymptotically smaller than the 
sample complexity of the best possible classical learning algorithm 
for the same problem.

Bernstein and Vazirani's algorithm is particularly striking because it
requires only a single query to the membership oracle to learn any 
concept in the problem class.  Only very special concept learning 
problems have quantum sample complexity 1 in this sense.  In \S2 we
explain that these are learning problems in what should be described
as ``Hadamard'' concept classes.  Other learning problems, like the
one solved by Grover's algorithm, have quantum sample complexity 
greater than 1, but one can ask how well a student can learn with a 
single query.  In \S3 we pose this problem of ``impatient learning'' 
precisely, and show that it is answered by the solution to a certain
geometric optimization problem.

With additional queries we should expect superior results.  In \S4 we 
show that the quantum computing technique of amplitude amplification 
[\Grover--\BHMT]
corresponds to querying also the other half of a minimally adequate 
teacher, the equivalence oracle.  Using an equivalence oracle we can 
define a general quantum learning algorithm, but without the use of 
some structure in---or symmetry of---the concept class, it is 
precisely Grover's algorithm, with the queries interpreted as being to 
the equivalence oracle, rather than to the membership oracle.  In \S5 
we review group algebras, in order to describe particular symmetries 
of concept classes.  These symmetries---{\it via\/} quantum fast 
transforms---allow equivalence queries to be combined with optimal 
impatient learning algorithms to achieve performance superior to use 
of equivalence queries alone.  In \S6 and \S7 we analyze the 
resulting quantum algorithms for concept classes with $\Z_N$ and 
$\Z_2^n$ symmetry, respectively.  We obtain efficient quantum
algorithms for two novel problems:  \battleship\ and \majority.

We conclude in \S8 with a discussion of the optimality of our quantum
algorithms, and their relevance to a pair of conjectured upper bounds
for the sample complexity of quantum learning algorithms.

\medskip
\noindent{\bf 2.  Formalization of quantum learning algorithms}

Bernstein and Vazirani's search problem is the task of identifying 
$a \in \Z_2^n$, given a `sophisticated' oracle that returns 
$a \cdot x$ mod 2 when queried about $x \in \Z_2^n$ [\noe].  From our 
point of view, it can also be interpreted as an instance of active 
learning with access to a membership oracle.  There is a family of 
concept classes ${\cal BV}^n$ for $0 < n \in \Z$, with
$$
{\cal BV}^n
 =
\bigl\{p_a : \Z_2^n \to \Z_2 \bigm| 
       p_a(x) = a\cdot x \hbox{\ mod\ }2 \hbox{\ for\ } a\in\Z_2^n
\bigr\},
$$
consisting of the concepts ``bit string with odd inner product with
$a$'' for $a\in\Z_2^n$.  Since the concept class ${\cal BV}^n$ is
parameterized by $a \in \Z_2^n$, identifying $a$ is equivalent to 
learning a target concept $p_a$ by querying a membership oracle.   
Classically this learning problem has sample complexity $\Omega(n)$.

In Bernstein and Vazirani's quantum algorithm for this problem, as
well as in all the quantum concept learning algorithms we consider in 
this paper, the ``data structure'' consists of a query ``register'' 
and a response ``register''---the Hilbert space of states is 
$\C^{|X|}\otimes \C^2$.  A membership oracle for target concept 
$\bar c$ acts {\it via\/} the unitary transformation $U_{\bar c}$ 
defined by linear extension from its action on the {\sl computational 
basis}, $\bigl\{|x,b\rangle \bigm| x \in X, b\in\Z_2\bigr\}$, namely 
$U_{\bar c}|x,b\rangle = |x,b+{\bar c}(x)\rangle$, where ``$+$'' 
denotes addition modulo 2.  Let 
$|-\rangle = \bigl(|0\rangle - |1\rangle\bigr)/\sqrt{2}$.  Then 
$U_{\bar c}|x\rangle|-\rangle = (-1)^{{\bar c}(x)}|x\rangle|-\rangle$.  
We will use this ``phase kickback'' trick [\CEMM] throughout, so we 
need only concentrate on the query register and, abusing notation 
slightly, write $U_{\bar c}|x\rangle = (-1)^{{\bar c}(x)}|x\rangle$. 

With this notation, Bernstein and Vazirani's algorithm is summarized 
by the equation:
$$
H^{\otimes n}U_{p_a}H^{\otimes n}|0\rangle = |a\rangle,     \eqno(2.1)
$$
where $H = \bigl({1\atop 1}{1 \atop -1}\bigr)/\sqrt{2}$ and 
$0\in\Z_2^n$.  That is, from an initial state $|0\rangle$, we apply
the {\sl Hadamard transform}, $H^{\otimes n}$; query the membership 
oracle; and apply the Hadamard transform again.  The result is the 
state $|a\rangle$, so a measurement in the computational basis 
identifies the target concept $p_a$ with probability 1.  The quantum 
sample complexity of this algorithm is 1, a substantial improvement 
over the classical sample complexity.

To understand why this algorithm works, notice that after the first
Hadamard transform in (2.1), the state of the query register is in an 
equal superposition of all possible queries:
$$
H^{\otimes n}|0\rangle 
 = {1\over\sqrt{2^n}}\sum_{x\in\sZ_2^n}|x\rangle.
$$
Such an equal superposition is the state before the initial query in 
each of the quantum algorithms we discuss in this paper.  Acting on 
this state by $U_{p_a}$ produces one of $|\Z_2^n| = 2^n$ possible 
vectors, according to the value of $a$.  Let $A_{\cal BV}$ be the 
matrix that has these vectors as columns.  In general we make the 
following definition.

\noindent\Definition.\ \ For any concept class ${\cal C}$ defined over
a set $X$, define the {\sl membership query matrix\/} $A_{\cal C}$ to 
be the $|X|\times |{\cal C}|$ matrix with $c^{\rm th}$ column 
$$
U_c {1\over\sqrt{|X|}}\sum_{x\in X}|x\rangle,
$$
for $c \in {\cal C}$.  In this paper we only consider concept classes 
for which there is a bijection between $X$ and ${\cal C}$; we call 
these {\sl matched\/} concept classes.  For matched concept classes, 
the membership query matrix is square.

For the Bernstein and Vazirani problem, the membership query matrix 
has entries 
$$
(A_{\cal BV})_{xa} = {(-1)^{x\cdot a}\over\sqrt{2^n}},
$$ 
which we recognize as the entries of $H^{\otimes n}$.  Thus the final
Hadamard transform in (2.1) acts as
$$
H^{\otimes n} (A_{\cal BV})_a 
 = (H^{\otimes n} A_{\cal BV})_a
 = (H^{\otimes n} H^{\otimes n})_a
 = (I)_a
 = |a\rangle,                                               \eqno(2.2)
$$
since $H = H^{-1}$.  That is, it inverts the query matrix.  Clearly,
then, the sample complexity of quantum learning in any concept class
with a unitary membership query matrix is 1.  Since such a membership 
query matrix is just a Hadamard matrix in the traditional sense (an 
orthogonal matrix with entries $\pm1$) [\Sylvester,\Hadamard], 
normalized by $\sqrt{|X|}$, we refer to such concept classes as 
{\sl Hadamard\/} concept classes.%
\sfootnote*{This nomenclature is motivated by van Dam's paper on a
quantum algorithm for the quadratic residue problem [\vanDam].}

Not all learning problems, of course, are this easy.  Grover's search
problem can also be interpreted as an instance of active concept 
learning with access to a membership oracle.  In this case there is a 
family $\{{\cal G}_N\}$ of concept classes, for $0 < N \in \Z$, with
$$
{\cal G}_N 
 = 
\bigl\{\delta_a : \Z_N \to \Z_2 \bigm| 
       \delta_a(x) = \delta_{ax} \hbox{\ for\ } a \in \Z_N
\bigr\},
$$
which consists of the concepts ``is the number $a$'' for $a \in \Z_N$.
The task is to identify $a$ given an oracle that returns $\delta_{ax}$
when queried about $x$.  Since the concept class ${\cal G}_N$ is
parameterized by $a \in \Z_N$, identifying $a$ is equivalent to 
identifying a target concept $\delta_a$.  Classically this learning 
problem has sample complexity $\Omega(N)$.  

Quantum mechanically, this oracle acts by a unitary matrix 
$U_{\delta_a}$, so the membership query matrix for this problem has 
entries
$$
(A_{\cal G})_{xa} 
 = {(-1)^{\delta_{xa}}\over\sqrt{N}}
 = {1\over\sqrt{N}}
   \bigl(NF^{\dagger}|0\rangle\langle0|F - 2I\bigr)_{xa},   \eqno(2.3)
$$
where $F$ is the $N$-dimensional discrete Fourier transform.  Clearly 
$A_{\cal G}$ is not unitary, so ${\cal G}_N$ is not a Hadamard 
concept class, and a single query does not suffice to learn a target
concept.  In fact, Bernstein and Vazirani [\BernsteinVazirani] showed
that (in our language) the sample complexity of Grover's learning 
problem is $\Omega(\sqrt{N})$.  Nevertheless, one might ask how well 
it is possible to do with a single query.  That is, if we can make any
unitary transformation (independent of $a$) after a single query, how 
do we maximize the probability that a measurement in the 
{\sl computational basis\/} $\bigl\{|x\rangle\bigm| x \in \Z_N\bigr\}$ 
returns $a$?  We give a general solution to this problem of 
{\sl impatient learning\/} in the next section, and then apply it to 
Grover's problem in \S5. 

\medskip
\noindent{\bf 3.  Impatient learning}

The column vectors of a membership query matrix---the possible states 
of the query register after a single equal superposition membership
query---form a special case of a general situation we can consider,
namely a quantum system whose state is one of a set of $0 < N \in \Z$ 
unit vectors $\bigl\{|v_i\rangle\bigm| i\in\Z_N\bigr\}$ in an
$N$-dimensional Hilbert space, ${\cal H}$.  
The task is to select a measurement to 
perform that will maximize the probability of correctly guessing which 
state the system was in before the measurement was made.  This is a
special case of the problem originally considered by Helstrom
[\Helstrom] and Kholevo [\Kholevo], {\sl quantum hypothesis testing}, 
namely identifying one from among a set of pure quantum states, no 
matter their provenance.

Recall that a von Neumann measurement [\vonNeumann] is defined by an 
orthogonal direct sum decomposition of the Hilbert space.  The 
measurement is {\sl complete\/} if  the summands are one-dimensional.  
Belavkin [\Belavkin] and Kennedy [\Kennedy] have shown that when the
$\{|v_i\rangle\}$ are linearly independent the optimal quantum 
measurement is, in fact, a complete von Neumann measurement.  Such a
measurement determines an orthonormal basis 
$\bigl\{|e_i\rangle\bigm| i\in\Z_N\bigl\}$, up to phases.  The 
probability that the system will be in state $|e_i\rangle$ after this 
measurement, given that it was in state $|v\rangle$ before the 
measurement, is $|\langle v|e_i\rangle|^2$.  If we assume that the 
system has been prepared in one of the states $\{|v_i\rangle\}$, 
chosen uniformly at random, then the quantity we want to maximize is
$$
\sum_{i=1}^N |\langle v_i| e_i\rangle|^2.                   \eqno(3.1)
$$
Necessary and sufficient criteria for solutions to this optimization
problem, in the more general case of arbitrary prior probabilities for
the $\{|v_i\rangle\}$, can be found in the early quantum hypothesis
testing literature [\Kholevo,\Belavkin,\Helstrombook].  In the 
following we provide a brief, geometrical derivation of such a 
criterion.

We can phrase this problem as a question about matrices:  If we choose
an isomorphism of Hilbert spaces, ${\cal H}\simeq \C^N$, then the list
$\bigl(|v_1\rangle,\ldots,|v_N\rangle\bigr)$ is identified with a 
square matrix $A \in M_N(\C)$.  Making an arbitrary complete 
measurement is equivalent to making an arbitrary unitary 
transformation, followed by a fixed complete measurement in, for 
example, the computational basis.  Thus we should consider the 
matrices $SA$, for $S\in U(N)$, where $U(N)$ denotes the unitary group.  
We write $A\sim B$ if $B = SA$ for some $S\in U(N)$.  Maximizing the 
quantity (3.1) is equivalent to maximizing the quantity
$$
\|d(B)\|^2                                                  \eqno(3.2)
$$
over the $U(N)$-orbit of $A$, $\{B \mid B\sim A\}$, where 
$d : M_N(\C) \to M_N(\C)$ is projection onto diagonal matrices and 
$\|\cdot\|$ is the $L^2$ (or Frobenius) norm.  In the following, when 
we speak of critical points of the function (3.2), it will be implicit
that the $U(N)$-orbit of $A$ is the domain.  We have the following 
characterization of the critical points:

\noindent\Proposition\ 3.1.  
{\sl The matrix $B$ is a critical point of $\|d(B)\|^2$ if and only if
$Bd(B)^{\dagger}$ is Hermitian,} \ie,
$$
Bd(B)^{\dagger} = d(B)B^{\dagger},
$$
{\sl where $^{\dagger}$ denotes the adjoint.}

\noindent{\sl Proof}.  Let ${\frak u}(N)$ denote the Lie algebra of 
$U(N)$, \ie, the set of skew-Hermitian matrices.  The criticality 
condition is that for all $\zeta\in{\frak u}(N)$,
$$
{{\rm d}\over{\rm d}t}\bigg|_{t=0} 
\bigl\|d\bigl((1+t\zeta)B\bigr)\bigr\|^2 = 0,
$$
which is true when ${\rm Re}\bigl(d(B)^{\dagger}d(\zeta B)\bigr) = 0$.  
But
$$
{\rm Re}\bigl(d(B)^{\dagger}d(\zeta B)\bigr) 
 = 
{\rm Re}\bigl({\rm tr}\bigl(\zeta Bd(B)^{\dagger}\bigr)\bigr),
$$
so the condition for $B$ to be critical is that $B d(B)^{\dagger}$ be
orthogonal to all skew-Hermitian matrices, with respect to the inner
product 
${\rm Re}\bigl({\rm tr}\bigl((\cdot)^{\dagger}(\cdot)\bigr)\bigr)$.  
This proves the proposition, since the orthogonal complement to the  
space of skew-Hermitian matrices is the  space of Hermitian matrices.
                                                      \hfill\endproof

This result seems to have been stated and proved (differently) first 
by Helstrom, in the more general setting of an arbitrary probability
distribution over the state vectors 
[\Helstrombook, Chap.~IV, eq.~(1.30)].  Since $\|d(B)\|^2$ is 
invariant under left multiplication by unitary diagonal matrices, we 
can restrict our attention to those critical points of (3.2) that have
nonnegative real entries on the diagonal.  Now the criticality 
condition reads
$$
Bd(B)=d(B)B^{\dagger}.                                      \eqno(3.3)
$$

We would like, however, an explicit solution to (3.3).  Consider the
Gram matrix of $A$, $G = A^\dagger A$, with components 
$G_{ij} = \langle v_i|v_j\rangle$.  $G$ is a positive semi-definite 
Hermitian matrix.  Let $\sqrt{G}$ denote the positive semi-definite
Hermitian square root of $G$.  By the polar decomposition of $A$, 
there is always a unitary matrix $S$ such that $\sqrt{G} = SA$, so it 
is natural to ask whether $\sqrt{G}$ is a critical point of (3.3).  
Proposition 3.1 shows that this is generally not the case.  More 
precisely, we have the following corollary:

\noindent\Corollary\ 3.2.  $\sqrt{G}$ {\sl is a critical point of\/}
(3.2) {\sl if and only if $\sqrt{G}$ commutes with its own diagonal.}

If the off-diagonal part of $\sqrt{G}$ is sufficiently general then
the conclusion of Corollary~3.2 will force the diagonal to be 
constant.  Although this is a strong condition in general, it is a 
very natural simplification [\YKLa,\Kholevo,\Helstrombook,\Belavkin].
We shall see that it occurs in many structured learning problems.
Moreover, having a constant diagonal is precisely the condition needed 
to go beyond impatient learning---which we will do in the next 
section.  So it is a case worth considering.

\noindent\Proposition\ 3.3.  {\sl Let $G$ be a positive semi-definite 
Hermitian matrix.  Let $\sqrt{G}$ denote the positive semi-definite 
Hermitian square root of $G$.  Assume the diagonal of $\sqrt{G}$ is 
constant.  Let ${\cal S}$ denote the set of matrices $B$ such that\/} 
$B\sim\sqrt{G}$ and $B$ {\sl has constant diagonal.  Then the maximum 
of $\|d(B)\|^2$ over $B\in {\cal S}$ occurs at $\sqrt{G}$.}

\noindent{\sl Proof}.  If $B$ has constant diagonal, then 
$\|d(B)\|^2 = |{\rm tr}(B)|^2/N$.  So it suffices to prove that 
$\sqrt{G}$ gives the maximum value of $|{\rm tr}(B)|^2$ over all 
$B\in{\cal S}$.  As in the proof of Proposition~3.1, the critical 
points occur when 
${\rm Re}\bigl({\rm tr}(B){\rm tr}(\zeta B)\bigr)=0$ for all 
$\zeta\in{\frak u}(N)$.  Writing 
${\rm tr}(B){\rm tr}(\zeta B) 
 = {\rm tr}\bigl(\zeta B\,{\rm tr}(B)\bigr)$, 
we see that the critical points are given by the condition that
$B\,{\rm tr}(B)$ is Hermitian.  Let $B_h=(B+B^{\dagger})/2$ and
$B_s=(B-B^{\dagger})/2$.  We want the skew-Hermitian part of 
$B\,{\rm tr}(B)$ to  vanish, thus 
$$
B_h{\rm tr}(B_s) + B_s{\rm tr}(B_h) =0.                     \eqno(3.4)
$$
The trace of (3.4) shows that either ${\rm tr}(B_h) = 0$ or
${\rm tr}(B_s) = 0$.  If both traces vanish then we get the minimum
possible value, $|{\rm tr}(B)|^2 = 0$.  If this is also the maximum
then $\sqrt{G}$ is forced to vanish since it is positive 
semi-definite, so the statement is true in this case.  If only one of 
the traces vanishes, the maximum occurs at a point where $B_s = 0$ or 
$B_h = 0$.  Since multiplication by $i$ is a symmetry of 
$|{\rm tr}(B)|^2$, we may assume $B_s = 0$.  Then $B$ is some square 
root of $G$.  The maximum of $|{\rm tr}(B)|^2$ will occur when one 
chooses the same sign for the square root of each eigenvalue, \eg, 
when $B=\sqrt{G}$.                                    \hfill\endproof

\noindent\Remark.  As we have noted above, both Proposition 3.1 and 
Proposition 3.3 have long been known in the context of quantum 
hypothesis testing.  Nevertheless, we have included our proofs of 
these results in order to emphasize the connection with a similar 
optimization problem:  These new proofs are inspired by the 
proof of the result that the minimum of the $L^2$ distance $\|B-I\|$ 
over the set $B\sim\sqrt{G}$ is given by $B=\sqrt{G}$, irrespective of 
any assumption about the diagonal.  In particular, for an arbitrary 
invertible matrix $A$, with polar decomposition $A=S^{-1}P$, where $S$ 
is unitary and $P$ is positive definite Hermitian, the closest point 
to $I$ in the $U(N)$-orbit of $A$ is $P$---this is the solution to the 
{\sl Procrustes Problem\/} [\procrustean].  Most 
recently, Eldar and Forney have noted that when the diagonal of 
$\sqrt{G}$ is constant, this solution to this optimization problem is 
also the solution to the optimization problem (3.1) that is the 
relevant one for quantum measurement [\EldarForney].

Thus we have the following quantum algorithm for a concept learning 
problem with membership query matrix $A_{\cal C}$:

\item{}{\bf Impatient Learning}
\parskip=0pt

\itemitem{\bf 1.}
Prepare the query register in the equal superposition state, 
$F^{\dagger}|0\rangle$, where $F$ is the $|X|$-dimensional discrete 
Fourier transform.  (Any unitary map taking $|0\rangle$ to the equal
superposition state works; in the case where $X = \Z_2^n$, 
$H^{\otimes n}$ can be applied.)

\itemitem{\bf 2.}
Query the membership oracle, obtaining as the state the 
$\bar c^{\rm th}$ column of $A_{\cal C}$, 
$U_{\bar c} F^{\dagger}|0\rangle$.

\itemitem{\bf 3.}
Apply a unitary transformation $S_{\cal C}$ such that 
$B_{\cal C} = S_{\cal C}A_{\cal C}$ satisfies (3.3).

\itemitem{\bf 4.}
Measure the resulting state 
$S_{\cal C}U_{\bar c} F^{\dagger}|0\rangle$ in the computational 
basis.

\parskip=10pt
\noindent As an immediate corollary of Proposition~3.1 we have:

\noindent\Theorem\ 3.4.  {\sl Impatient Learning succeeds with 
probability $|(B_{\cal C})_{\bar c\bar c}|^2$ and is optimal among
single query quantum algorithms that begin with an equal superposition
over membership queries.}  

As we saw in (2.2), for $A_{\cal BV}$, 
$B_{\cal BV} = H^{\otimes n} A_{\cal BV} = I$ maximizes (3.3), so the 
Bernstein-Vazirani algorithm {\sl is\/} Impatient Learning, and 
succeeds with probability 1 for every target concept.  Furthermore, as 
we will see in \S5, for Grover's problem, (3.3) is maximized by
$B_{\cal G} = 
 \bigl(2F^{\dagger}|0\rangle\langle0|F - I\bigr)A_{\cal G}$.  
Using (2.3) it is then easy to compute that the diagonal entries of 
$B_{\cal G}$ are all $(3-4/N)/\sqrt{N}$, so Impatient Learning 
succeeds with asymptotic probability $9/N$ as $N \to \infty$.  
Theorem~3.4 says that this is the best we can do using only a 
single membership query.  Although it is certainly an improvement over 
the success probability $1/N$ of random guessing, Impatient Learning 
is far from satisfactory for this problem. 

\medskip
\noindent{\bf 4.  Beyond Impatient Learning}

\nobreak
In fact, for most concept learning problems a single membership query
simply does not provide enough information to learn the target concept
with probability close to 1.  A specific target concept $\bar c$ 
defines a subspace of the Hilbert space $\C^{|X|}$, namely 
span$\{|\bar c\rangle\}$ (recall that there is a bijection between
$X$ and ${\cal C}$), however, so we can apply one of the few general
quantum algorithm techniques---amplitude amplification 
[\Grover--\BHMT].
This technique, 
invented by Brassard and H{\o}yer [\BrassardHoyer] as a generalization 
of Grover's algorithm [\Grover], can be described in terms of 
concepts:

\noindent A{\eightpoint MPLITUDE} A{\eightpoint MPLIFICATION} 
([\BHMT], Theorem~2).
{\sl Let $\chi$ be a concept over $X$; let ${\cal H}_1$ denote the
subspace of $\C^{|X|}$ spanned by the vectors labeled by the elements
in the extension of $\chi$, $\chi^{-1}(1)$; and let $\Pi$ denote the
projection $\C^{|X|} \to {\cal H}_1$.  For any unitary transformation 
$W$ of $\C^{|X|}$, let $p(W) = \bigl|\Pi W|0\rangle\bigr|^2$ be the 
probability with which the state $W|0\rangle$ is measured to be in the 
subspace ${\cal H}_1$.  As long as $p(W) > 0$, we can set 
$\sin^2\theta = p(W)$ for $0 < \theta \le \pi/2$.  In this case, 
repeatedly applying the unitary transformation
$WU_{\delta_0}W^{\dagger}U_{\chi}$ amplifies the probability of 
measuring the state to be in the subspace ${\cal H}_1$.  More 
precisely,
$$
p\bigl((WU_{\delta_0}W^{\dagger}U_{\chi})^m W\bigr)
 \ge \max\{1-p(W),p(W)\},
$$
where $m = \big\lfloor{\pi\over4\theta} - {1\over2}\big\rceil$, the 
nearest integer to ${\pi\over4\theta} - {1\over2}$.}

After step {\bf 3} of Impatient Learning, the state is 
$S_{\cal C}U_{\bar c}F^{\dagger}|0\rangle$, where $S_{\cal C}$ was 
chosen to maximize 
$\sum_{\bar c} 
\bigl|\langle\bar c|S_{\cal C}U_{\bar c}F^{\dagger}|0\rangle\bigr|^2$.  
Thus, letting $\chi = \delta_{\bar c}$, 
${\cal H}_1 = {\rm span}\{|\bar c\rangle\}$ and 
$W_{\cal C} = S_{\cal C}U_{\bar c}F^{\dagger}$, applying Amplitude 
Amplification gives a new quantum algorithm:

\item{}{\bf Amplified Impatient Learning}
\parskip=0pt

\itemitem{\bf 1.}
Prepare the query register in the equal superposition state, 
$F^{\dagger}|0\rangle$, where $F$ is the $|X|$-dimensional discrete 
Fourier transform.  (Any unitary map taking $|0\rangle$ to the equal
superposition state works; in the case where $X = \Z_2^n$, 
$H^{\otimes n}$ can be applied.)

\itemitem{\bf 2.}
Query the membership oracle, obtaining as the state the 
$\bar c^{\rm th}$ column of $A_{\cal C}$, 
$U_{\bar c} F^{\dagger}|0\rangle$.

\itemitem{\bf 3.}
Apply an impatient learning transform $S_{\cal C}$, producing 
$S_{\cal C}U_{\bar c}F^{\dagger}|0\rangle = W_{\cal C}|0\rangle$.

\itemitem{\bf 4.}
Apply $W_{\cal C}U_{\delta_0}W_{\cal C}^{\dagger}U_{\delta_{\bar c}}$ 
$m$ times, where 
$m = \big\lfloor{\pi\over4\theta} - {1\over2}\big\rceil$, and 
$\sin\theta = \bigl|\langle\bar c|W_{\cal C}|0\rangle\bigr| = 
|(B_{\cal C})_{\bar c\bar c}|$, with $0 < \theta \le {\pi\over2}$.

\itemitem{\bf 5.}
Measure the resulting state in the computational basis.

\parskip=10pt
\noindent As a consequence of Theorem~3.4 and Amplitude Amplification 
we have:

\noindent\Theorem\ 4.1.  {\sl For problems with $B_{\cal C}$ having 
constant diagonal element $s$, Amplified Impatient Learning succeeds 
with probability at least $\hbox{\rm max}\{1 - s^2,s^2\}$.  Since each 
of $W_{\cal C}$ and $W_{\cal C}^{\dagger}$ includes calls to the 
membership oracle {\it via} $U_{\bar c}$, Amplified Impatient Learning 
has sample complexity $2m + 1$, \ie, $O(1/s)$.}

Notice, however, that the algorithm uses more than membership queries.  
The operation $U_{\delta_{\bar c}}$ in step {\bf 4} is the action of 
an {\sl equivalence\/} oracle responding to a queried {\sl concept\/} 
(rather than a concept argument).  Roughly speaking, the Impatient 
Learning part of this algorithm maximizes the amplitude for the target 
concept $\bar c$ after a single membership query; then an equivalence 
oracle is queried about the correctness of this concept.  Thus 
Amplified Impatient Learning uses {\sl both\/} oracles comprising the 
minimally adequate teacher defined in \S1, making $2m + 1$ membership 
queries and $m$ equivalence queries.

At the risk of confusing the membership and equivalence oracles, we
can apply Amplified Impatient Learning to Grover's problem.  As we 
noted at the end of \S3 and as we will compute in \S5, for 
$A_{\cal G}$ the post-membership query transform is 
$S_{\cal G} = 2F^{\dagger}|0\rangle\langle0|F - I 
            = F^{\dagger}U_{\delta_0}F$.
So for this problem, 
$$
W_{\cal G} 
  = S_{\cal G}U_{\bar c}F^{\dagger} 
  = F^{\dagger}U_{\delta_0}FU_{\delta_{\bar c}}F^{\dagger},
$$
where this use of $U_{\delta_{\bar c}}$ is a query to the membership
oracle.  The iterated transformation is
$$
\eqalign{
W_{\cal G}U_{\delta_0}W_{\cal G}^{\dagger}U_{\delta_{\bar c}}
 &=
(F^{\dagger}U_{\delta_0}FU_{\delta_{\bar c}}F^{\dagger})
U_{\delta_0}
(FU_{\delta_{\bar c}}F^{\dagger}U_{\delta_0}F)
U_{\delta_{\bar c}}                                                \cr
 &=
F^{\dagger}U_{\delta_0}FU_{\delta_{\bar c}}\cdot
F^{\dagger}U_{\delta_0}FU_{\delta_{\bar c}}\cdot
F^{\dagger}U_{\delta_0}FU_{\delta_{\bar c}}                        \cr
 &=
(F^{\dagger}U_{\delta_0}FU_{\delta_{\bar c}})^3,                   \cr
}
$$
where the $U_{\delta_{\bar c}}$ in the first expression is the 
operation of the equivalence oracle but the distinction between the 
two kinds of oracles is ignored in the last expression.  The complete
algorithm is then
$$
(F^{\dagger}U_{\delta_0}FU_{\delta_{\bar c}})^{3m}
 F^{\dagger}U_{\delta_0}FU_{\delta_{\bar c}}F^{\dagger}|0\rangle
 = 
(F^{\dagger}U_{\delta_0}FU_{\delta_{\bar c}})^{3m+1}F^{\dagger}|0\rangle,
$$
where $m = \big\lfloor{\pi\over4\theta} - {1\over2}\big\rceil$, and 
$\theta = \arcsin |(B_{\cal G})_{\bar c\bar c}| 
        = \arcsin \bigl((3 - 4/N)/\sqrt{N}\bigr)$.
Thus $m \sim {\pi\over4}\sqrt{N}/3$ so the interated transformation is 
applied ${\pi\over4}\sqrt{N}$ times, asymptotically.  This is, in 
fact, exactly Grover's algorithm [\Grover], although one usually sees
it factored differently (and with $F$ and $F^{\dagger}$ replaced by 
$H^{\otimes n}$).

\medskip
\noindent{\bf 5.  Concept classes with group symmetry}

The sets $\Z_2^n$ and $\Z_N$, over which the Bernstein-Vazirani and 
Grover concept classes are defined, are abelian groups under 
componentwise addition modulo 2 and addition modulo $N$, respectively.
In each case the Hilbert space $\C^G$ becomes a ring, with
multiplication law defined by linear extension from
$$
|x\rangle * |y\rangle = |x+y\rangle {\rm\ for\ } x,y \in G,
$$
where $G$ is $\Z_2^n$ or $\Z_N$.

\noindent\Definition.  The {\sl group algebra\/} of $G$ is the Hilbert
space $\C^G$ (often written $\C[G]$), equipped with this ring 
structure.

\noindent The regular representation of the group algebra is the
map
$$
\C^G \ni |v\rangle \longmapsto L_v \in M_{|G|}(\C),
$$
where $L_v$ is left multiplication by $|v\rangle$, a linear map on 
$\C^G$, hence a $|G|\times|G|$ complex matrix in the computational 
basis.  We will identify the group algebra with its image in this 
representation.

For $G = \Z_2$, 
$\C^2 \ni |v\rangle = \alpha|0\rangle + \beta|1\rangle$ is a general 
element of the group algebra.  From the definition,
$$
\eqalign{
|v\rangle * |0\rangle &= \alpha|0\rangle + \beta|1\rangle          \cr
|v\rangle * |1\rangle &= \alpha|1\rangle + \beta|0\rangle,         \cr
}
$$
so 
$$
L_v = \pmatrix{\alpha &  \beta \cr
                \beta & \alpha \cr
              } 
    = \alpha I + \beta X
    = \alpha X^0 + \beta X^1,                               \eqno(5.1)
$$
where $X = \bigl({0\atop1}{1\atop0}\bigr) = L_1$ is the usual Pauli 
matrix.  More generally, we have:

\noindent\Proposition\ 5.1.  {\sl The group algebra of $\Z_2^n$ 
consists of $2^n\times2^n$ dimensional matrices of the form
$$
L_v = \sum_{x\in\sZ_2^n} v_x X^x,
$$
for
$$
|v\rangle = \sum_{x\in\sZ_2^n} v_x|x\rangle 
\in \C^{\sZ_2^n} = (\C^2)^{\otimes n}.
$$
}({\sl In the expression for $L_v$, $x\in\Z_2^n$ is a multi-index, 
\ie, 
$X^x = X^{x_1\ldots x_n} = X^{x_1}\otimes\cdots\otimes X^{x_n}$.})  
{\sl $|v\rangle$ is the first column of $L_v$; $L_v$ is symmetric and
has constant diagonal.  $L_v$ is diagonalized by the Hadamard 
transform.}

\noindent{\sl Proof}.  By induction on $n$.  That the Hadamard 
transform diagonalizes the elements of the $\Z_2^n$ group algebra 
follows from the familiar Pauli matrix identity $Z = HXH$, where 
$Z = \bigl({1\atop0}{0\atop-1}\bigr)$.                  \hfill\endproof

The group $\Z_N$ is generated by the element 1, and for $y \in \Z_N$,
$$
L_1 : |y\rangle \mapsto |1+y\rangle
\quad\Longrightarrow\quad
L_1 = \pmatrix{  0  &      &\cdots&     &   1  \cr
                 1  &   0  &      &     &      \cr
                    &\ddots&\ddots&     &\vdots\cr
                    &      &\ddots&   0 &       \cr
                    &      &      &   1 &   0   \cr
              }.
$$
Analogously to Proposition~5.1, for $\Z_N$ we have:

\noindent\Proposition\ 5.2.  {\sl The group algebra of $\Z_N$ consists 
of $N\times N$ dimensional matrices of the form
$$
L_v = \sum_{x\in\sZ_N}v_x L_1^x,
$$
where
$$
|v\rangle = \sum_{x\in\sZ_N}v_x|x\rangle \in \C^{\sZ_N} = \C^N.
$$
$|v\rangle$ is the first column of $L_v$; $L_v$ need not be symmetric, 
but it has constant diagonal.  $L_v$ is diagonalized by the 
$N$-dimensional discrete Fourier transform.}

\noindent{\sl Proof}.  That the Fourier transform diagonalizes the 
elements of the $\Z_N$ group algebra follows from the fact that 
$FL_1F^{\dagger} = {\rm diag}(1,\omega,\omega^2,\ldots,\omega^{N-1})$,
where $\omega = e^{2\pi i/N}$.                         \hfill\endproof

Propositions 5.1 and 5.2 allow us to characterize useful symmetries of 
concept classes:

\noindent\Proposition\ 5.3.  {\sl Let ${\cal C}$ be a matched concept
class over an abelian group $G$.  Then $A_{\cal C}$ is in the group 
algebra of $G$ if and only if it commutes with the action of $G$, \ie,
$$
L_g A_{\cal C} = A_{\cal C} L_g, \quad\forall g\in G.       \eqno(5.2)
$$
In components\/} (5.2) {\sl becomes
$$
(A_{\cal C})_{x+g,c+g} = (A_{\cal C})_{xc}, \quad\forall g\in G;
                                                            \eqno(5.3)
$$
equivalently $c(x)$ is a function of $c-x$.}

In Grover's problem, $A_{\cal G}$ satisfies (5.2) and (5.3) for 
$G = \Z_N$ (and for $G = \Z_2^n$, when $N = 2^n$).  Thus this 
membership query matrix belongs to the group algebra of $\Z_N$ (and of
$\Z_2^n$, when $N = 2^n$) and is diagonalized by $F$ (and by 
$H^{\otimes n}$, when $N = 2^n$).  It is, furthermore, a real 
symmetric matrix.  The following proposition explains how to compute 
the optimal transformation $S$ required for Impatient Learning and 
Amplified Impatient Learning in this case.

\noindent\Proposition\ 5.4.  {\sl Let $A$ be a real, symmetric matrix
in the group algebra of $\Z_2^n$} ({\sl or $\Z_N$\/}).  {\sl Using the 
Spectral Theorem, define $|A|$ by 
$$
|A|v = |\lambda|v,
$$
for each eigenvector-eigenvalue pair $(v,\lambda)$ of $A$.  Then $|A|$
is also an element of the same group algebra.  Moreover, the maximum
value of $\|d(B)\|^2$ over matrices $B\sim A$ with constant diagonal
occurs at $|A|$.}

\noindent{\sl Proof}.  Since $A$ is real and symmetric, $A$ is a 
square root of its Gram matrix.  Conjugation by the appropriate
transform ($H^{\otimes n}$ or $F$) diagonalizes $A$ so $|A|$, having
the same eigenvectors, is also an element of the same group algebra as
$A$.  Moreover, $|A|$ is the positive semi-definite square root of the 
Gram matrix.  Thus the result follows from Proposition 3.3.
                                                       \hfill\endproof

Proposition 5.4, applied to a symmetric membership query matrix 
$A_{\cal C}$ satisfying the conditions of Propostion 5.3, implies that
an optimal unitary transformation $S_{\cal C}$ in the Impatient 
Learning and Amplified Impatient Learning algorithms satisfies
$$
|A_{\cal C}| = S_{\cal C}A_{\cal C}.                        \eqno(5.4)
$$
When $A_{\cal C}$ is nonsingular, $S_{\cal C}$ is unique and (5.4) 
implies that
$$
S_{\cal C} = |A_{\cal C}|A_{\cal C}^{-1} =: {\rm sign}(A_{\cal C}),
$$
where 
$$
{\rm sign}(A)v = {\rm sign}(\lambda)v = {\lambda\over|\lambda|}v,
$$
for all eigenvector-eigenvalue pairs $(v,\lambda)$ of $A$.  

To compute $S_{\cal G}$ for Grover's concept class we diagonalize 
$A_{\cal G}$:
$$
F{1\over\sqrt{N}}
 \bigl(NF^{\dagger}|0\rangle\langle0|F - 2I
 \bigr)F^{\dagger}
 =
{1\over\sqrt{N}}\bigl(N|0\rangle\langle0| - 2I\bigr).
$$
This implies that 
$$
S_{\cal G}
   = {\rm sign}(A_{\cal G}) 
   = F^{\dagger}{\rm diag}(1,-1,\ldots,-1)F        
   = F^{\dagger}\bigl(2|0\rangle\langle0| - I\bigr)F
   = -F^{\dagger}U_{\delta_0}F, 
$$
which is the promised expression for $S_{\cal G}$ that we quoted in 
\S3 and \S4.

\medskip
\noindent{\bf 6.  Learning problems with cyclic symmetry}

Although recognizing Grover's algorithm as an instance of Amplified 
Impatient Learning perhaps contributes to a better understanding of 
this basic quantum algorithm, we would like to apply the general 
formalism developed in the preceding sections to derive new quantum 
algorithms.  So in this section we consider some new problems with 
cyclic symmetry.

According to Proposition~5.3, any learning problem with a transitive
$\Z_N$ action has the property that the oracle response $c(x)$ depends 
only on the difference $c-x$ mod $N$.  Thus we may write 
$c(x) = \phi(c-x)$ for some function $\phi : \Z_N \to \Z_2$, whence 
the membership query matrix $A_{\cal C}$ is
$$
A_{\cal C} = {1\over\sqrt{N}} \sum_{k\in\sZ_N} (-1)^{\phi(k)} L_1^k.
                                                            \eqno(6.1)
$$
By Proposition~5.2, $A_{\cal C}$ is diagonalized by the Fourier 
transform; hence its eigenvalues are
$$
\lambda_j 
 = 
{1\over\sqrt{N}} \sum_{k\in\sZ_N} (-1)^{\phi(k)} \omega^{jk},
                                                            \eqno(6.2)
$$
for $j \in\Z_N$.  By Proposition~5.4, the relevant quantity for 
Impatient Learning is the size of the diagonal elements of 
$|A_{\cal C}|$.  This matrix is in the $\Z_N$ group algebra, and hence 
has constant diagonal.  Furthermore, the diagonal element $s$ is just
the average of the eigenvalues of $|A_{\cal C}|$, \ie, the average of 
the absolute values of the eigenvalues of $A_{\cal C}$.  By 
Theorem~4.1, therefore, Amplified Impatient Learning requires $O(1/s)$ 
queries.

Consider a special class of cyclically symmetric problems, which we 
call \battleship, after the Milton Bradley game with the same name.  
Let $0\leq r < N/2$.  For any $a,x \in \Z_N$, set
$$
b_a(x) = \cases{1 & if $a-x \equiv -r,\ldots,r$ mod $N$;           \cr
                0 & otherwise.                                     \cr
}
$$
$d = 2r+1$ is the {\sl length\/} of the battleship, \ie, $d$ counts 
the number of $x \in \Z_N$ that satisfy $b_a(x) = 1$ for any fixed 
$a$.  

It turns out that the behavior of \battleship\ problems depends on the 
relative size of $d$ with respect to $N$.  Thus we consider two 
separate subfamilies of \battleship:  For the problem \smallship, we 
fix the value of $d$ and let $N$ be arbitrary.  For the problem 
\bigship, we again let $N$ be arbitrary, but fix the ratio 
$\alpha\in (0,1/2)$ of $d$ to $N$.  That is, we take 
$d=\lfloor\alpha N\rceil$.  

\noindent\Theorem\ 6.1.  {\sl For any fixed $d$, Amplified Impatient 
Learning solves {\rm\smallship}\ with $O(\sqrt{N})$ queries, which 
is optimal to within a constant factor.  When applied to 
{\rm\bigship}, however, Amplified Impatient Learning requires 
$\Omega(\sqrt{N}/\log N)$ queries, which is far from optimal.}

\noindent{\sl Proof}.  The eigenvalues of $A_{\cal BS}$ for the 
\battleship\ concept class with parameters $N$ and $r$ are
$$
\lambda_j = {1\over\sqrt{N}} 
            \Bigl( -\sum_{k=-r}^r \omega^{jk} +
                    \sum_{k=r+1}^{N-r-1} \omega^{jk} 
            \Bigr).
$$
In particular, this gives
$$
\lambda_0 = {1\over\sqrt{N}} (N-2d), 
$$
while for $j>0$,
$$
\lambda_j = -{2\over\sqrt{N}} {\sin(\pi jd/N)\over\sin(\pi j/N)}. 
                                                           \eqno(6.3)
$$

First, consider the case of \smallship\ for fixed $d$.  Since the
expression $\sin(\pi j/N)$ in the denominator of (6.3) is bounded 
above in absolute value by $1$, it follows that
$$
s\sqrt{N}
 \ge 
{2\over N} \sum_{j=1}^{N-1} \bigl|\sin{\pi jd\over N}\bigr|.
$$
As $N$ tends to infinity, the right hand side approaches the constant
value 2$\int_0^1 |\sin d\pi x | {\rm d}x$.  We conclude that 
$s = \Omega(1/\sqrt{N})$ and hence Amplified Impatient Learning has
query complexity $O(\sqrt{N})$.  To see that this is optimal to
within a constant factor, note that for each $a$ there are only $d$ 
values of $x$ for which $b_a(x) = 1$.  Thus any classical learning 
algorithm which uses only the membership oracle requires $\Omega(N/d)$ 
queries.  Note also that the equivalence oracle can be simulated 
using exactly 2 calls to the membership oracle, since $c = b_a$ if and
only if $c(a+r) = 1$ and $c(a+r+1) = 0$.  Hence the equivalence oracle 
is unnecessary, and the results of Servedio and Gortler 
[\ServedioGortler] imply that we can achieve at most a quadratic 
speedup over the classical algorithm.  Thus $\Omega(\sqrt{N/d})$ 
quantum queries are required.  Since $d$ is constant, we see that for
\smallship\ Amplified Impatient Learning is optimal up to a constant 
factor.

Second, consider \bigship\ for fixed $\alpha$.  In this case we claim 
that $s = O\bigl((\log N)/\sqrt{N}\bigr)$.  To see this, note first 
that $\lambda_0/N$ is $O(1/\sqrt{N})$.  Bounding each of the sines in 
the numerator of (6.3) by $1$, we find that for $j > 0$, $\lambda_j/N$ 
is bounded above in absolute value:
$$
{1\over N} |\lambda_j| \leq {1\over N\sqrt{N}} \csc {\pi j\over N}.
$$
It follows that $\lambda_1/N$ and $\lambda_{n-1}/N$ are 
$O(1/\sqrt{N})$, while the remaining sum
$$
{1\over N}\sum_{j=2}^{N-2} |\lambda_j| 
 \leq 
{1\over N\sqrt{N}} \sum_{j=2}^{N-2}\csc {\pi j\over N}
 \leq
{1\over \sqrt{N}} \int_{1/N}^{1-1/N}\!\!\csc{\pi x} \, {\rm d}x 
 = 
O\Bigl({\log N\over \sqrt{N}}\Bigr).
$$
Thus the number of steps required by Amplified Impatient Learning is
$\Omega(\sqrt{N}/\log N)$.  To see that this is not an optimal 
algorithm, consider using Grover's algorithm to return some $x$ for 
which $b_a(x) = 1$.  This requires $O(\sqrt{N/d})$ quantum queries
[\BBHT], and narrows the possible answer space to a set of size $d$.  
A classical binary search, requiring $\log d$ further (classical) 
queries can now be used to identify the answer $a$ uniquely.  This 
alternative algorithm solves \bigship\ with only 
$O(\sqrt{N/d} + \log d) = O(\sqrt{1/\alpha} + \log\alpha N) 
                        = O(\log N)$ 
queries, far fewer than the $\Omega(\sqrt{N}/\log N)$ required by 
Amplified Impatient Learning.                          \hfill\endproof 

\medskip
\noindent{\bf 7.  The M{\eightpoint\bf AJORITY} problem}

The other group algebra introduced in \S5 is that of $\Z_2^n$.  In 
this section we study a novel problem, \majority, that has this
symmetry.  Fix a positive integer $n$, and for each $a\in\Z_2^n$  
define a function $m_a: \Z_2^n \to \Z_2$ by
$$
m_a(x) = \cases{1 & if ${\rm wt}(a-x) \le n/2$;                   \cr    
                0 & otherwise.                                    \cr
               }
$$
That is, $m_a(x)=1$ when the bit strings $a$ and $x$ agree in at
least as many bits as they disagree.  The \majority\ concept class 
${\cal MAJ}^n$ is defined to be the set of all functions $m_a$, 
where $a$ is any element of $\Z_2^n$.  It is easy to see that any
classical learning algorithm requires at least $n$ queries.  We can do
better quantum mechanically: 

\noindent\Theorem\ 7.1.  {\sl Amplified Impatient Learning solves 
{\rm\majority}\ with $O(\sqrt{n})$ quantum queries, given access to 
both the membership oracle and the equivalence oracle.}

If $n$ is an odd integer, the membership query matrix for 
${\cal MAJ}^n$ contains in its upper left-hand corner (rows and 
columns labeled $0\ldots0$ to $1\ldots1$ top-to-bottom and 
left-to-right, respectively) a $2^{n-1} \times 2^{n-1}$ submatrix 
proportional to the membership query matrix for ${\cal MAJ}^{n-1}$.  
Furthermore, if $a,b\in\Z_2^n$ are complementary bit strings then 
$m_a(x) = 1-m_b(x)$ for all $x$, and hence the column in the 
membership query matrix corresponding to $a$ equals the negative of 
the column corresponding to $b$.  It follows that if one can learn a
concept from ${\cal MAJ}^{n-1}$, then one can learn a concept from 
${\cal MAJ}^n$ with one additional membership query.  {\sl Thus, in 
what follows, we will assume that $n$ is an even integer.}

For learning problems with $\Z_2^n$ symmetry we have the following 
analogues of (6.1) and (6.2):  Since $c(x) = \phi(c-x)$ for some
$\phi : \Z_2^n \to \Z_2$, the membership query matrix has the form
$$
A_{\cal C} 
 = 
{1\over\sqrt{2^{n}}} \sum_{b\in\sZ_2^{n}} (-1)^{\phi(b)} X^b.
                                                            \eqno(7.1)
$$
By Proposition~5.1, $A_{\cal C}$ is diagonalized by the Hadamard 
transform; hence its eigenvalues are
$$
\lambda_c 
 = 
{1\over\sqrt{2^{n}}}\sum_{b\in\sZ_2^{n}} 
 (-1)^{\phi(b)}(-1)^{b\cdot c},                             \eqno(7.2)
$$
for $c\in\Z_2^n$.  With these preliminaries in place, we can prove
Theorem~7.1:

\noindent{\sl Proof}.  For the concept class ${\cal MAJ}^n$, 
$\phi(b) = \Theta\bigl({n\over2}-{\rm wt}(b)\bigr)$, where the 
Heaviside function $\Theta(z) = 1$ if $z \ge 0$; and vanishes 
otherwise.  It is easy to see in (7.2) that for this problem the value 
of $\lambda_c$ depends only on $k = {\rm wt}(c)$.  Thus, for 
$k\in\{0,\dots,n\}$, we may set $\lambda_{n,k} = \lambda_c$, where 
$c\in\Z_2^n$ is any bit string of weight $k$.  To calculate 
$\lambda_{n,k}$, we consider the string $c=0^{n-k}1^k$, which has 
weight $k$.  For any $b\in\Z_2^{n}$, let $s$ denote the number of $1$s 
in the first $n-k$ bits of $b$, and let $r$ denote the number of $1$s 
in the remaining $k$ bits of $b$.  Then $b\cdot c = r$, and 
$\phi(b) = \Theta\bigl({n\over2}-(r+s)\bigr)$.  Since the number of 
bit strings $b$ with given values for $r$ and $s$ is
${n-k\choose s}{k\choose r}$, we have
$$
\lambda_{n,k} 
 = 
\sum_{r,s} {n-k\choose s}{k\choose r} (-1)^{\Theta({n\over2}-(r+s))} 
           (-1)^r.
$$
Using standard combinatorial techniques, this sum simplifies to give
$$
\lambda_{n,k}
 =
\cases{-{(-1)^{k/2}\over\sqrt{2^{n}}} {n\choose n/2} 
       {1\cdot3\cdots(k-1)\over(n-1)\cdot(n-3)\cdots(n-k+1)}
       & for $k$ even;                                             \cr
       \lambda_{n,k-1} & for $k$ odd.                              \cr
      }
$$
It follows that for any even number $n$, the eigenvalue of smallest 
absolute value is the middle eigenvalue $\lambda_{n,n/2}$, which 
is given by
$$
|\lambda_{n,n/2}| 
 = 
\cases{{1\over\sqrt{2^{n}}}{n/2\choose n/4}
       & if $n \equiv 0$ mod 4;                                    \cr 
       {2\over\sqrt{2^{n}}}{(n-2)/2\choose(n-2)/4}
       & if $n \equiv 2$ mod 4.                                    \cr
      }
$$
Since, by Stirling's formula, each of these expressions is asymptotic 
to $1/\sqrt{n}$, we find that the average $s$ of the absolute values 
of the eigenvalues of $A_{{\cal MAJ}^n}$ is $\Omega(1/\sqrt{n})$.  
Thus the quantum query complexity of \majority\ is  
$O(1/s) = O(\sqrt{n})$, as claimed.                    \hfill\endproof

\medskip
\noindent{\bf 8.  Conclusion}

\nobreak
In this paper we have derived a general technique---Amplified 
Impatient Learning---for quantum concept learning from a minimally 
adequate teacher.  We applied it to two novel problems:  \battleship\
and \majority, that like the problems of Bernstein-Vazirani and 
Grover, can be recognized as concept learning problems.  Amplified
Impatient Learning solves \smallship\ with $O(\sqrt{N})$ queries, an
improvement over the $\Omega(N)$ queries required classically.  For
\bigship, Amplified Impatient Learning is not so good, but we gave an
alternative quantum algorithm with sample complexity $O(\log N)$.  
Finally, Amplified Impatient Learning solves \majority\ with 
$O(\sqrt{n})$ quantum queries, again an improvement over the 
$\Omega(n)$ required classically.

Quantum algorithms for concept learning were first considered by
Bshouty and Jackson [\BshoutyJackson], who analyzed the traditional 
DNF learning problem [\Valiant].  Subsequently, Servedio and Gortler 
proved some general lower bounds on the quantum sample complexity of 
learning from any membership oracle [\ServedioGortler].  Their 
results, together with algorithms derived in this paper, motivate us 
to make a pair of conjectures about general {\sl upper\/} bounds on 
the quantum sample complexity of learning from a membership oracle:

\noindent\Conjecture\ 1.  For any family of concept classes 
$\{{\cal C}_i\}$ with $|{\cal C}_i| \to \infty$, there exists a 
quantum learning algorithm with membership oracle query complexity 
$O(\sqrt{|{\cal C}_i|})$.

Our quantum algorithm for \smallship\ (which specializes to Grover's
algorithm when $d = 1$) saturates this bound; the idea is that these
minimally structured search problems are concept learning problems 
that are as difficult as any of the same size.  As we noted in the
proof of Theorem~6.1, the calls to the equivalence oracle in this 
problem can be replaced by calls to the membership oracle, so our 
results are consistent with Conjecture~1.

The difficulty of concept learning problems depends on more than the
number of concepts $|{\cal C}|$ among which the target concept lies,
however; it also depends on how similar distinct concepts are.  
Servedio and Gortler express their lower bounds in terms of a quantity
$\gamma_{\cal C}$ that measures this similarity:  For any 
${\cal C}' \subset {\cal C}$, define 
${\cal C}'_{x,b} = \{c\in{\cal C}' \mid c(x) = b\}$.  Then
$$
\gamma_{\cal C} 
 = 
\min_{{\cal C}'\subset{\cal C},\,|{\cal C}'|\ge2}
\max_{x\in X}\min_{b\in\sZ_2} {|{\cal C}'_{x,b}|\over|{\cal C}|}.
$$   
$\gamma_{\cal C}$ is small if there is a large subset ${\cal C}'$ from 
which the response to any query to the membership oracle rules out 
only a small fraction of the concepts.  In this case we expect the 
concept class to be difficult to learn.  (Since ${\cal C}'$ can 
contain only 3 concepts, from which a query might eliminate only 1, 
$\gamma_{\cal C}$ cannot be greater than $1/3$.)

\noindent\Conjecture\ 2.  For any family of concept classes
$\{{\cal C}_i\}$ with $|{\cal C}_i| \to \infty$, there exists a 
quantum learning algorithm with membership oracle query complexity 
$O\bigl({\log|{\cal C}_i|\over\sqrt{\gamma}}\bigr)$.

The problems \majority, \smallship, and \bigship\ studied in this
paper provide examples of learning problems that satisfy the
bounds given in the above conjectures.  For the \battleship\ problem,
one calculates that $\gamma_{\cal BS} = \min\{d/N,1/3\}$.
Thus, for \smallship, whose quantum sample complexity is $O(\sqrt{N})$, 
Conjecture~1 is sharp, while Conjecture~2 provides the weaker bound of 
$O(\sqrt{N} \log N)$.  For \bigship, the situation
is reversed:  Conjecture~2 provides a sharp upper bound of $O(\log N)$,
while Conjecture~1 gives the weaker upper bound of $O(\sqrt{N})$. For
\majority, whose quantum sample complexity is at most
$\log N$, it is easy to see that Conjecture~2 holds by using the
fact that $\gamma \leq 1/3$.

\medskip
\noindent{\bf Acknowledgements}

We thank Zeph Landau, Jeff Remmel and Ronald de Wolf for useful 
discussions.  This work has been partially supported by the National 
Science Foundation (NSF) under grant ECS-0202087, and by the National
Security Agency (NSA) and Advanced Research and Development Activity 
(ARDA) under Army Research Office (ARO) grant number DAAD19-01-1-0520.

\medskip
\global\setbox1=\hbox{[00]\enspace}
\parindent=\wd1

\noindent{\bf References}
\bigskip

\parskip=0pt
\item{[\Shor]}
\shor,
``Algorithms for quantum computation:  discrete logarithms and 
  factoring'',
in S. Goldwasser, ed.,
{\sl Proceedings of the 35th Symposium on Foundations of Computer 
Science}, Santa Fe, NM, 20--22 November 1994
(Los Alamitos, CA:  IEEE 1994) 124--134;\hfb
\shor,
``Polynomial-time algorithms for prime factorization and discrete 
  logarithms on a quantum computer'',
{\tt quant-ph/9508027};
\SIAMJC\ {\bf 26} (1997) 1484--1509.

\item{[\Coppersmith]}
D. Coppersmith,
``An approximate Fourier transform useful in quantum factoring'',
IBM T. J. Watson Research Report RC 19642 (1994).

\item{[\Hoyer]}
P. H{\o}yer,
``Efficient quantum transforms'',
{\tt quant-ph/9702028}.

\item{[\Klappenecker]}
A. Klappenecker,
``Wavelets and wavelet packets on quantum computers'',
{\tt quant-ph/9909014};
in
M.~A.~Unser, A.~Aldroubi, A.~F.~Laine, eds., 
{\sl Wavelet Applications in Signal and Image Processing VII}, 
Denver, CO, 19--23 July 1999, 
{\sl SPIE Proceedings\/} {\bf 3813}
(Bellingham, WA:  SPIE 1999) 703--713.

\item{[\KlappeneckerRoetteler]}
A. Klappenecker and M. R\"otteler,
``Discrete cosine transforms on quantum computers'',
{\tt quant-ph/0111038};
in
S. Loncaric and H. Babic, eds.,
{\sl Proceedings of the 2nd International Symposium on Image and Signal
     Processing and Analysis}, 
Pula, Croatia, 19--21 June 2001
(Los Alamitos, CA:  IEEE 2001) 464--468.

\item{[\Freedman]}
M. H. Freedman, 
``Poly-locality in quantum computing'', 
{\tt quant-ph/0001077};
\FoCM\ {\bf 2} (2002) 145--154.

\item{[\Grover]}
\grover,
``A fast quantum mechanical algorithm for database search'',
in {\sl Proceedings of the 28th Annual ACM Symposium on 
  the Theory of Computing},
Philadelphia, PA, 22--24 May 1996 
(New York:  ACM 1996) 212--219;\hfb
\grover, 
``Quantum mechanics helps in searching for a needle in a haystack'', 
{\tt quant-ph/9706033};
\PRL\  {\bf 79} (1997) 325--328.

\item{[\BrassardHoyer]}
G. Brassard and P. H{\o}yer,
``An exact quantum polynomial-time algorithm for Simon's problem'',
{\tt quant-ph/9704027};
{\sl Proceedings of 5th Israeli Symposium on Theory of
     Computing and Systems}, Ramat-Gan, Israel 17--19 June 1997 
(Los Alamitos, CA:  IEEE 1997) 12--23.

\item{[\GroverB]}
\grover,
``A framework for fast quantum mechanical algorithms'',
{\tt quant-ph/ 9711043};
in
{\sl Proceedings of the 30th Annual ACM Symposium on Theory of Computing},
Dallas, TX, 23--26 May 1998
(New York:  ACM 1998) 53--62.

\item{[\BHT]}
G. Brassard, P. H{\o}yer and A. Tapp,
``Quantum counting'',
{\tt quant-ph/9805082};
{\sl Proceedings of the 25th International Colloquium on Automata,
Languages, and Programming}, {\AA}lborg, Denmark, 13--17 July 1998,  
{\sl Lecture Notes in Computer Science\/} {\bf 1443}
(Berlin:  Springer-Verlag 1998) 820--831.

\item{[\BHMT]}
G. Brassard, P. H{\o}yer, M. Mosca and A. Tapp,
``Quantum amplitude amplification and estimation'',
{\tt quant-ph/0005055};
in
S. J. Lomonaco, Jr.\ and H. E. Brandt, eds.,
{\sl Quantum Computation and Information}, 
{\sl Contemporary Mathematics\/} {\bf 305}
(Providence, RI:  AMS 2002) 53--74. 

\item{[\BernsteinVazirani]}
\bv,
``Quantum complexity theory'',
in {\sl Proceedings of the 25th ACM Symposium on Theory of Computing},
San Diego, CA, 16--18 May 1993
(New York:  ACM Press 1993) 11--20;\hfb
\bv,
``Quantum complexity theory'',
\SIAMJC\ {\bf 26} (1997) 1411--1473.

\item{[\machinelearning]}
For the computer science perspective see, \eg, 
T. M. Mitchell,
{\sl Machine Learning\/}
(San Francisco:  McGraw-Hill 1997);\hfb
for a recent mathematical perspective see
F. Cucker and S. Smale, 
``On the mathematical foundations of learning'',
\BAMS\ {\bf 39} (2002) 1--49.

\item{[\Angluinsurvey]}
See, \eg,
D. Angluin,
``Computational learning theory:  survey and selected bibliography'',
in
{\sl Proceedings of the 24th Annual ACM Symposium on Theory
     of Computing}, Victoria, British Columbia, Canada, 4--6 May 1992
(New York:  ACM 1992) 351--369.

\item{[\Angluin]}
D. Angluin, 
``Queries and concept learning'',
\ML\ {\bf 2} (1988) 319--342.

\item{[\GareyJohnson]}
See, \eg,
M. R. Garey and D. S. Johnson,
{\sl Computers and Intractability:  A Guide to the Theory of 
     NP-Completeness\/}
(New York:  W. H. Freeman 1979).

\item{[\noe]}
\dajm, 
``Sophisticated quantum search without entanglement'',
{\tt quant-ph/\hfb 0007070}; 
\PRL\ {\bf 85} (2000) 2014--2017.

\item{[\CEMM]}
R. Cleve, A. Ekert, C. Macchiavello and M. Mosca,
``Quantum algorithms revisited'',
{\tt quant-ph/9708016};
\PRSLA\ {\bf 454} (1998) 339--354.

\item{[\Sylvester]}
J. J. Sylvester,
``Thoughts on inverse orthogonal matrices, simultaneous 
  sign-succes\-sions, and tesselated pavements in two or more colours,
  with applications to Newton's rule, ornamental tile-work, and the
  theory of numbers'',
\PM\ ser.\ IV {\bf 34} (1867) 461--475.

\item{[\Hadamard]}
M. J. Hadamard,
``{\it R\'esolution d'une question relative aux d\'eterminants\/}'',
\BSM\ {\bf 17} (1893) 240--246.

\item{[\vanDam]}
W. van Dam,
``Quantum algorithms for weighing matrices and quadratic residues'',
{\tt quant-ph/0008059};
\Alg\ (2002) OF1--OF16.

\item{[\Helstrom]}
C. W. Helstrom,
``Detection theory and quantum mechanics'',
\IC\ {\bf 10} (1967) 254--291.

\item{[\Kholevo]}
A. S. Kholevo,
``Quantum statistical decision theory'',
\JMA\ {\bf 3} (1973) 337--394.

\item{[\vonNeumann]}
\vonneumann,
{\it Mathematische Grundlagen der Quantenmechanik\/}
(Berlin:  Spring\-er-Verlag 1932); 
transl.\ by R. T. Beyer as
{\sl Mathematical Foundations of Quantum Mechanics\/}
(Princeton:  Princeton University Press 1955).

\item{[\Belavkin]}
V. P. Belavkin,
``Optimal multiple quantum statistical hypothesis testing'',
\St\ {\bf 1} (1975) 315--345.

\item{[\Kennedy]}
R. S. Kennedy,
``On the optimal receiver for the $M$-ary pure state problem'',
{\sl MIT Res.\ Lab.\ Electron.\ Quart.\ Prog.\ Rep.}\ {\bf 110}
(15 July 1973) 142--146;\hfb
see also [\Helstrombook], Appendix to Chap.~IV.

\item{[\Helstrombook]}
C. W. Helstrom,
{\sl Quantum Detection and Estimation Theory\/}
(New York:  Academic 1976).

\item{[\YKLa]}
H. P. Yuen and R. S. Kennedy,
``On optimal quantum receivers for digital signal detection'',
\PIEEE\ {\bf 58} (1970) 1770--1773.

\item{[\procrustean]}
See, \eg,
R. A. Horn and C. R. Johnson, 
{\sl Matrix Analysis\/} 
(Cambridge:  Cambridge University Press 1985), 
p.~432, Theorem 7.4.9.

\item{[\EldarForney]}
Y. C. Eldar and G. D. Forney, Jr.,
``On quantum detection and the square-root measurement'',
{\tt quant-ph/0001532};
\IEEETIT\ {\bf 47} (2001) 858--872.

\item{[\ServedioGortler]}
R. A. Servedio and S. J. Gortler,
``Quantum versus classical learnability'',
in 
{\sl Proceedings of the 16th Annual IEEE Conference on Computational 
     Complexity}, Chicago, IL, 18--21 June 2001
(Los Alamitos, CA:  IEEE 2001) 138--148. 

\item{[\BBHT]}
M. Boyer, G. Brassard, P. H{\o}yer and A. Tapp,
``Tight bounds on quantum searching'',
{\tt quant-ph/9605034};
\ForP\ {\bf 46} (1998) 493--506.

\item{[\BshoutyJackson]}
N. H. Bshouty and J. C. Jackson,
``Learning DNF over the uniform distribution using a quantum example
  oracle'',
\SIAMJC\ {\bf 28} (1999) 1136--1153.

\item{[\Valiant]}
L. G. Valiant,
``A theory of the learnable'',
\CACM\ {\bf 27} (1984) 1134--1142.

\bye